\documentclass[
 reprint,
 superscriptaddress,
 amsmath,amssymb,
 aps,
]{revtex4-2}

\usepackage{graphicx}
\usepackage{dcolumn}
\usepackage{bm}

\usepackage{hyperref}
\hypersetup{
    colorlinks=true,
    linkcolor=blue,
    citecolor=blue,
    urlcolor=blue
}

\begin{document}


\title{Monte-Carlo study of Compositional Heterogeneity in Multicomponent Cluster Crystals}

\author{Roshan Maharana}
\email{rm2293@cam.ac.uk}
\affiliation{Institute of Physics, Chinese Academy of Sciences, Beijing 100190, China}
\affiliation{Yusuf Hamied Department of Chemistry, University of Cambridge, Lensfield Road, Cambridge CB2 1EW, United Kingdom}

\author{Daan Frenkel}
\email{df246@cam.ac.uk}
\affiliation{Yusuf Hamied Department of Chemistry, University of Cambridge, Lensfield Road, Cambridge CB2 1EW, United Kingdom}

\author{Jure Dobnikar}
\email{jd489@cam.ac.uk}
\affiliation{Institute of Physics, Chinese Academy of Sciences, Beijing 100190, China}
\affiliation{Wenzhou Institute of the University of Chinese Academy of Sciences, Wenzhou, Zhejiang 325011, PR China.}
\affiliation{Faculty for Mathematics and Physics, University of Ljubljana, Slovenia}
\date{\today}
\begin{abstract}
Soft (sub)micron-sized particles with bounded interactions can form cluster crystals, periodic structures in which multiple particles occupy the same lattice site. 
While the thermodynamics of monodisperse cluster crystals is well understood, less is known about how compositional disorder affects their stability. 
Using Monte Carlo simulations and density functional theory we show that binary cluster crystals undergo a density driven transition from a homogeneous mixed state to a heterogeneous ``alloy" like solid in which lattice sites spontaneously differentiate into populations with distinct compositions and occupancies while preserving the underlying crystal symmetry. 
The transition is accompanied by a sharp increase in the equilibrium lattice site density and by increased compositional fluctuations, but we see no evidence for macroscopic phase separation. We demonstrate that this transition is governed by competition between clustering and demixing instabilities and derive a simple scaling law for the demixing density as a function of temperature, composition, and particle size mismatch, in quantitative agreement with simulation.
\end{abstract}

\maketitle
Ultrasoft, bounded, repulsive interactions arise naturally in many soft matter systems, including polymer chains~\cite{PhysRevLett.85.2522, doi:10.1021/acs.macromol.1c02610}, dendrimers~\cite{B601916C,10.1063/1.1689292}, microgels~\cite{bergman2018new}, and DNA-based nanostructures~\cite{stiakakis2021self}. 
Such interactions are classified as either $Q^{+}$ or cluster forming $Q^{\pm}$ depending on whether the Fourier transform of the pair potential remains positive or becomes negative at finite wavevector~\cite{PhysRevE.63.031206}. 
Systems with $Q^{\pm}$ interactions, such as the generalized exponential model $\exp[-(r/\sigma)^m]$ with $m>2$, form cluster crystals at high densities in which multiple particles occupy the same lattice site~\cite{PhysRevLett.96.045701,PhysRevLett.99.235702,10.1063/1.2738064,mladek2008multiple}.
The existence of cluster crystal phases has been demonstrated in both simulations and experiments in systems ranging from amphiphilic dendrimers and DNA-based nanostructures to supersolid droplet crystals in dipole blockaded quantum gases~\cite{PhysRevLett.100.028301,doi:10.1021/acsnano.3c06083,doi:10.1021/acs.jpclett.1c01916,10.1063/1.4950953,PhysRevLett.109.228301,doi:10.1021/acs.jpcb.7b11662,stiakakis2021self,PhysRevLett.105.135301}.
Cluster crystals exhibit unusual properties including iso-structural transitions between different occupancies, reentrant melting~\cite{PhysRevLett.105.245701,10.1063/1.4723869,PhysRevE.86.042501,10.1063/1.4901302,van2009cluster}, unconventional elastic responses, multiple phonon branches associated with occupancy fluctuations, defect mediated relaxation, and particle hopping between lattice sites~\cite{neuhaus2011phonon,PhysRevB.92.184103,10.1063/5.0073624,montes2013hopping}. Their nonequilibrium behavior, including nucleation, growth, melting, and response to external driving, has also been explored extensively~\cite{leitold2016nucleation,PhysRevE.99.042140,PhysRevLett.107.068302,delfau2017active}.

\begin{figure*}[t!]
\centering
\includegraphics[width=2.05\columnwidth]{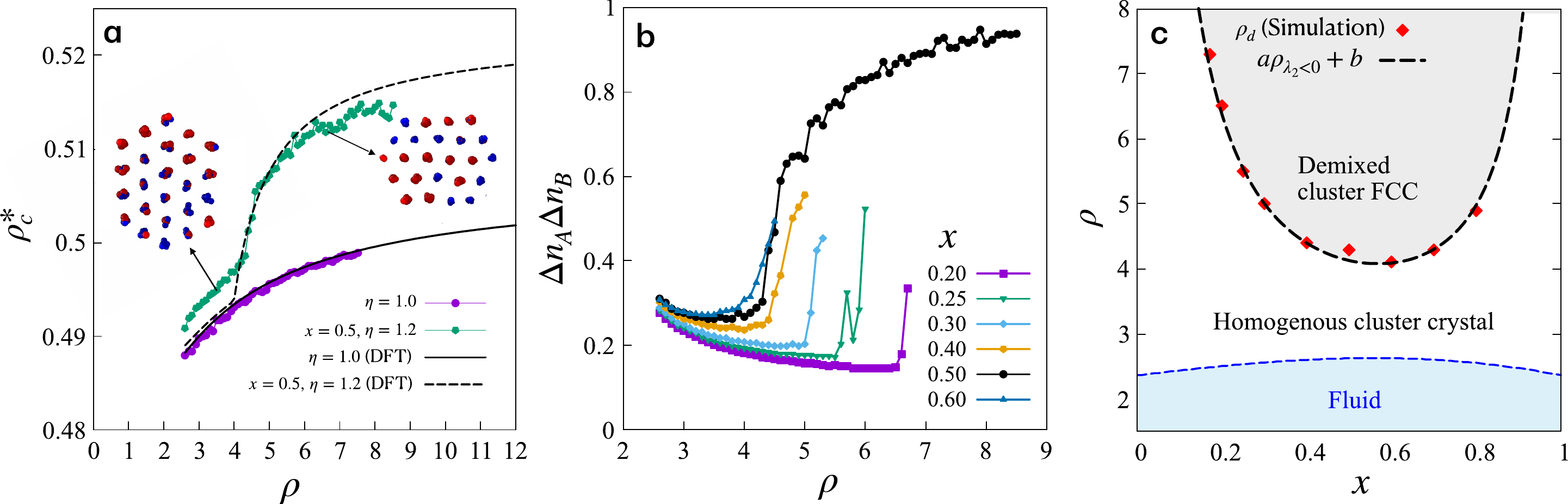}
\caption{\label{fig:rho_rhoc}
\textbf{(a)} Equilibrium lattice site density $\rho_c^*$ as a function of particle density $\rho$ for monodisperse and bidisperse ($\eta=1.2, x=0.5$) systems at $T=0.3$. The solid and dashed black lines correspond to DFT results. (\textbf{Inset:} A single $(111)$ layer of the cluster FCC crystal before and after the transition, where color blue: $\textbf{A}-$particles and red: $\textbf{B}-$particles)
\textbf{(b)} Compositional heterogeneity parameter $\Delta n_A \Delta n_B$ as a function of $\rho$ for different global compositions $x$. For each $x$, a sharp jump occurs at $\rho=\rho_d$, signaling the onset of cluster compositional heterogeneity.
\textbf{(c)} Composition–density phase diagram for a bidisperse FCC cluster crystal with $\eta=1.2$ at $T=0.3$. The blue dashed line corresponds to the $\lambda_1$ instability line of the homogeneous fluid. The red points denote the numerically determined values of $\rho_d$ for different $x$. The dashed black line has a form $\rho_d = a \rho_{\lambda_2<0}+b$, where $a,b$ are fitting parameters and  $\rho_{\lambda_2<0}$ provided in Eq.~\eqref{eq:rho_d}.
}
\label{fig:fig_1}
\end{figure*}

While the thermodynamics of monodisperse cluster crystals is well understood~\cite{PhysRevLett.99.235702,likos2008cluster,acs.jpcb.6b05471}, many experimental realizations are intrinsically multicomponent, with particles differing in size, shape, or interaction strength.  
Amphiphilic assemblies, micelles, and block copolymer aggregates provide important examples where morphology and internal composition depend sensitively on molecular architecture and thermodynamic conditions~\cite{polym14214702,C0PY00379D}. 
In particular, studies of micelles formed by binary surfactant mixtures report two possible scenarios, namely the formation of mixed micelles or demixed aggregates depending on the interaction parameters and composition~\cite{doi:10.1021/jp801216e,doi:10.1021/la503458n,doi:10.1021/acs.analchem.7b00339}. 
On the theoretical side, most studies of binary ultrasoft systems have focused on the fluid phase~\cite{10.1063/1.4926469,10.1063/5.0209814}, while other works showed that mixtures in which one component belongs to the $Q^{\pm}$ class and the other to the $Q^{+}$ class exhibit demixing and complex phase behavior even at small concentrations of non cluster forming particles~\cite{overduin2009phase}. 
Classical DFT has recently been used to study the formation and phase behaviour of 50:50 binary cluster crystals with asymmetric interactions and equal particle sizes~\cite{Tscharnutter2026}.
It is, however, not known how, size disparity between the constituents of a two component mixture influences the equilibrium properties of binary cluster crystals. 
These questions are of interest because mixed crystals have  compositional degrees of freedom at each lattice site, allowing mixed occupancies and new forms of symmetry breaking and instability. 
In this work we investigate the stability of multicomponent ultrasoft cluster crystals, focusing primarily on bidisperse systems. We consider a three dimensional binary mixture interacting via the generalized exponential model of order four (GEM-4):
$\phi_{\mu\nu}(r)=\epsilon_{\mu\nu}\exp[-(r/\sigma_{\mu\nu})^4],$
where $\mu,\nu\in \{A,B\}$ denote particle species, $\epsilon_{\mu\nu}$ the interaction strength, and $\sigma_{\mu\nu}$ the interaction range. 
This bounded repulsive potential belongs to the $Q^{\pm}$ class and forms cluster crystals at high densities. 
We study mixtures with compositions $x_A\equiv x$ and $x_B=1-x$. 
We use reduced units  with $\epsilon_{\mu\nu}=1$ and $x_{\mu}\sigma_{\mu\mu}^3=1$, while bidispersity is introduced through the size ratio $\eta=\sigma_{BB}/\sigma_{AA}$ with $\sigma_{AB}=(\sigma_{AA}+\sigma_{BB})/2$. 
We performed Canonical Monte Carlo simulations at fixed particle number $N$, volume $V$, temperature $T$, and composition. 
In all cases we used  periodic boundary conditions. 
Systems are initialized in crystalline configurations (fcc or bcc) with multiple particles per lattice site and equilibrated using Monte Carlo moves that include displacement moves, particle swaps and hops between lattice sites. 
The lattice site density of cluster crystals is defined as  $\rho_c=N_c/V$, where $N_c$ is the number of lattice sites and $V$ the volume of the crystal. 
We can define the lattice site chemical potential as $\mu_c=(\partial F/\partial N_c)_{N,V,T}$.
At equilibrium,  $\mu_c$ is equal to zero, because at equilibrium the free energy of the system is (to linear order) independent $\rho_c^*$. 
For equilibrated configurations at  different state points $(\eta,x,\rho,T)$, we measured the average lattice site occupancy $\overline{n}$, species resolved occupancies $\overline{n_A}$ and $\overline{n_B}$, and orientational order parameters $q_4$ and $q_6$.

We first determined the equilibrium lattice site density $\rho_c^*$ from the condition $\mu_c=0$ for monodisperse GEM-4 cluster crystals. 
As expected, $\rho_c^*$ increases monotonically with the total density $\rho$, reflecting the increase in mean lattice site occupancy while preserving the underlying FCC structure. 
The corresponding occupancy distribution remains unimodal and narrow, indicating that all lattice sites are statistically equivalent and the crystal is compositionally and structurally homogeneous. 
In contrast, bidisperse systems exhibit qualitatively different behavior. Figure~\ref{fig:fig_1}\textbf{(a)} shows that the $\rho_c^*$–$\rho$ curve initially follows a monotonic growth but jumps to a different branch beyond a threshold density $\rho_d$, which depends on temperature, global composition, and particle size ratio. 
These two branches correspond to distinct linear growth regimes of lattice site occupancy, signalling the emergence of two characteristic length scales associated with the lattice spacing. 
A qualitative understanding of this transition can be obtained by considering the free energy minimisation of a multicomponent cluster crystal with respect to the local occupancies of $A$ and $B$ particles as well as their spatial localisation parameters. 
Within this framework the free energy can be written as
\begin{equation}
    \begin{aligned}
        F&
= 
k_BT\sum_{i=1}^{N_c}\sum_{\mu}
n_i^{\mu}(\ln \left(n_i^{\mu}C_{\mu})-5/2\right)\\
&+
\sum_i
\sum_{\mu,\nu}
\Big[
n^i_{\mu}n^i_{\nu}
-\delta_{\mu\nu}n^i_{\mu}
\Big]
U_{\mu\nu}
+
\sum_{i\neq j}
\sum_{\mu,\nu}
n^i_{\mu}n^j_{\nu}
V_{\mu\nu},
    \end{aligned}
    \label{eq:F}
\end{equation}
where $n^i_\mu$ denotes the number of particles of species $\mu$ occupying lattice site $i$, while $U_{\mu\nu}$ and $V_{\mu\nu}$ represent the intra site and inter site interaction contributions, respectively. 
Details of this free energy functional are provided in the Supplemental materials. 
In Fig.~\ref{fig:fig_1} (dashed black line), we notice that the two observed branches appear naturally in during minimisation of $F$, where the first branch correspond to a homogeneous occupancy and composition state whereas the second branch correspond to appearance of two different kinds of clusters distributed across sites. 
In Eq.~\eqref{eq:F}, the ideal contribution favours mixing of the different species on each lattice site, whereas the intra site interaction term promotes demixing on the same site by lowering the free energy. 
The inter site interaction term penalizes large scale segregation and therefore suppresses macroscopic phase separation. 
The competition between these contributions allows for the possibility of compositionally distinct clusters while preserving the overall lattice structure. 

To identify the microscopic origin of the transition, we examine the species resolved lattice site occupancies $n^i_A$ and $n^i_B$. 
The homogeneity of the crystal is quantified using the fluctuations in total and species resolved occupancies ($\Delta n, \Delta n_A, \Delta n_B$). 
In particular, we consider the product $\Delta n_A \Delta n_B$, which is $0$ for a completely homogeneous state and approaches $1$ for a completely demixed state (see Supplemental materials for details). 
At low densities the system forms a homogeneous mixed cluster crystal in which both species occupy all lattice sites, with unimodal occupancy and composition distributions centered at the global composition $x_\mu$. 
For $\rho<\rho_d$, the system becomes increasingly homogeneous with increasing density, as indicated by the decrease of $\Delta n_A \Delta n_B$ with $\rho$ in Fig.~\ref{fig:fig_1}\textbf{(b)}.
At $\rho_d$, the system undergoes a sharp transition to a heterogeneous state, marked by a sudden increase in $\Delta n_A \Delta n_B$ and the emergence of bimodal composition distributions corresponding to $A$-rich and $B$-rich clusters. 
However, the system does not undergo complete demixing, as quantified by $\Delta n_A \Delta n_B<1$, although this quantity approaches unity with increasing density.  
Importantly, the $A$-rich and $B$-rich clusters remain spatially mixed within the same lattice rather than separating into distinct crystal domains, while the underlying FCC lattice symmetry remains unchanged. The average spatial width of the clusters, pressure and chemical potential vary smoothly with density without any signature of the transition. 
The demixing transition is followed by an increase in local occupancy fluctuations as particles redistribute between A-rich and B-rich sites so as to equalize their local energies. 
For identical interaction amplitudes ($\epsilon_{\alpha\beta}=1$), this occupancy imbalance between B- and A-rich sites grows linearly with density, $\overline{n_B^{i} -n_A^{i}} \propto \overline{n} =\rho/\rho_c^*$. 
The ratio of number of $A$-rich to $B$-rich lattice sites is given by, $N_c^A/N_c^B = \Delta n_B/\Delta n_A$.

Next, we analyze the stability of a homogeneous fluid by separating the free energy response into contributions from \emph{density} and \emph{composition} modulations. 
This analysis predicts a simple scaling relation for the demixing density. 
For an $M$-component system, fluctuations $\rho_\alpha=\rho_\alpha^0+\delta\rho_\alpha$ lead to a quadratic change in free energy governed by a stability matrix $H_{\alpha\beta}(q)$ in Fourier space. 
One eigenmode of $\textbf{H}$ corresponds to a \emph{density modulation} that drives cluster formation at the wavevector $q^*$ where the Fourier transform of the interaction potential is most negative, while the remaining $M-1$ eigenmodes correspond to \emph{composition fluctuations} that can lead to demixing.
In the present work, we focus on the bidisperse case ($M=2$), where the two eigenmodes correspond to total density ($\delta\rho_A+\delta\rho_B$) and composition difference ($\delta\rho_A-\delta\rho_B$). 
We find that the density mode becomes unstable first, $\lambda_1(q_1^*)<0$, at a lower density than the composition mode for all temperatures when $\eta<1.25$, indicating that cluster crystal formation occurs before demixing sets in.
\begin{figure}[t!]
\centering
\includegraphics[width=0.99\columnwidth]{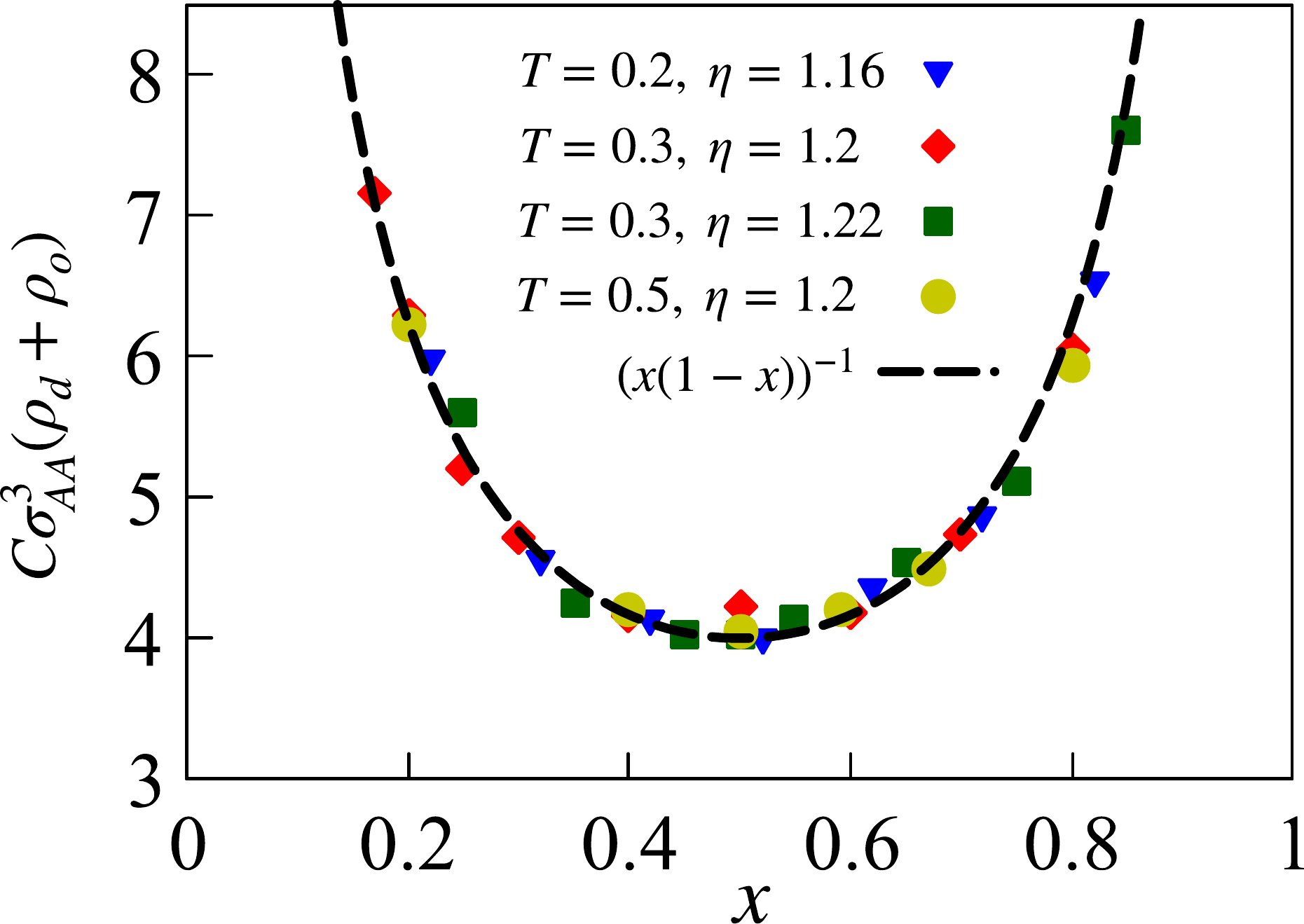}
\caption{\label{fig:fig_2}
Composition scaling of the demixing density. 
The scaled quantity $C\sigma_{AA}^3(\rho_d+\rho_o)$ follows the predicted $1/[x(1-x)]$ behavior for different $T$ and $\eta$. $C, \rho_o$ are fitting parameters.
}
\end{figure}
\begin{figure*}[t!]
\centering
\includegraphics[width=2.0\columnwidth]{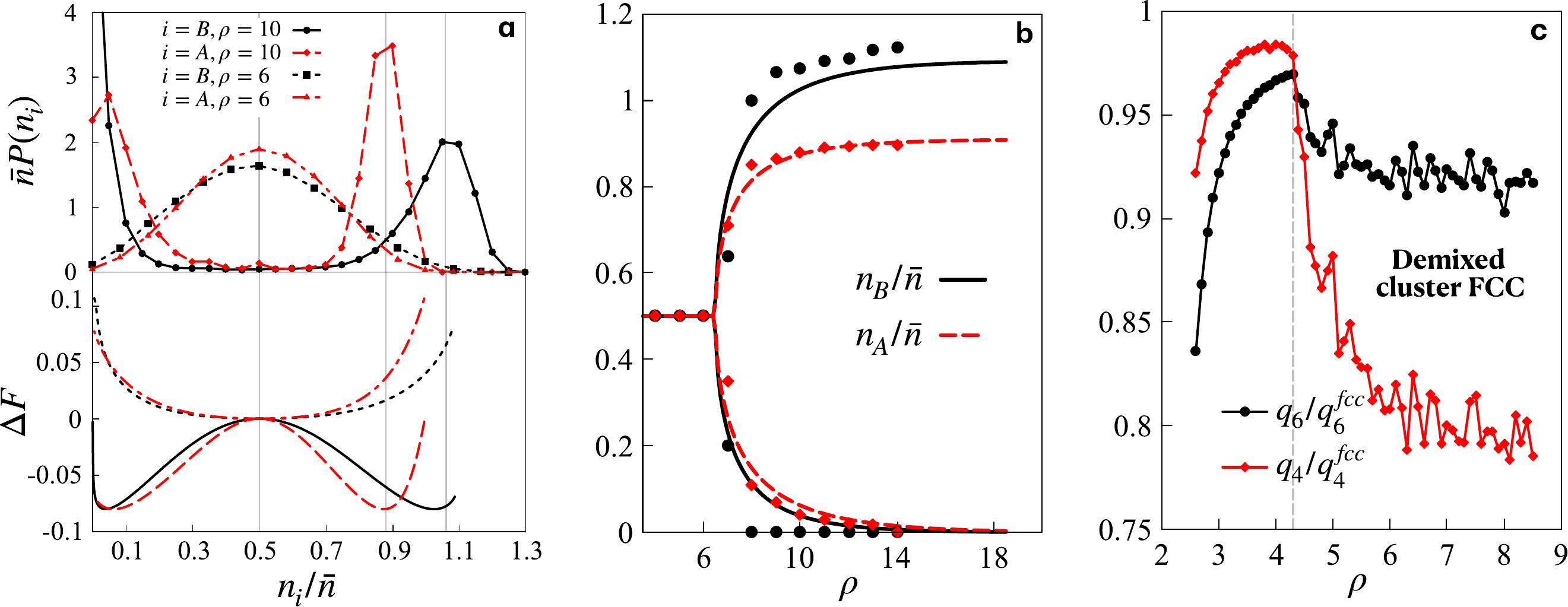}
\caption{\label{fig:fig_3}
\textbf{(a)} Top: Distribution of $A$ and $B$ particle occupancies in individual clusters, scaled by the average total cluster occupancy $\overline{n}$, before and after the demixing transition ($\rho_d \approx 6.5$ for $T=0.5, \eta=1.2, x=0.5$, and $\rho_c=0.5$). \\
Bottom: Corresponding free energy landscape $\Delta F$ obtained by minimising Eq.~\eqref{eq:F} while constraining the local occupancies to fixed values of $n_A/\overline{n}$ and $n_B/\overline{n}$.\\
\textbf{(b)} Average cluster occupancies of $A$ and $B$ particles, scaled by $\overline{n}$, as a function of $\rho$ for the same global composition, showing the splitting of occupancies for $\rho \ge \rho_d$.
\textbf{(c)} Orientational order parameters $q_6$ and $q_4$, scaled by their values for a perfect FCC lattice ($q_6^{fcc} \approx 0.57452$, $q_4^{fcc} \approx 0.19094$), as a function of density. Simulation results are shown for $\eta=1.2$, $x=0.5$, $T=0.3$ at the equilibrium lattice site densities $\rho_c^*$.
}
\end{figure*}
However, this analysis provides the correct scaling for the demixing density, since the local density within clusters can exceed the demixing threshold even when the global density is lower. For the bidisperse system this yields
\begin{equation}
\rho_{\lambda_2<0} =
\frac{k_BT}{x(1-x)\left(\sum_{\mu\nu}\tilde \phi_{\mu\nu}(q_2^*)
-\sum_{\mu}\tilde \phi_{\mu\mu}(q_2^*)
\right)},
\label{eq:rho_d}
\end{equation}
where $\tilde{\phi}_{\mu \nu}(q_2^*) = \sigma_{\mu \nu }^3\tilde{\phi}(\sigma_{\mu \nu }q_2^*)$.
We therefore assume that the demixing density in the cluster crystal phase satisfies $\rho_d \propto \rho_{\lambda_2<0}$, which agrees well with the numerical results (Fig.~\ref{fig:fig_1}\textbf{(c)}). 
As shown in Fig.~\ref{fig:fig_2}, the scaled quantity $C\sigma_{AA}^3(\rho_d+\rho_o)$ follows the predicted $1/[x(1-x)]$ dependence for different $T$ and $\eta$, with $C,\rho_o$ are fitting parameters.
This argument extends naturally to multicomponent systems. 
While the clustering density depends primarily on the average interaction and is therefore weakly affected by the number of components, the demixing instability depends on differences between self and cross interactions. 
As the number of components increases (at equal composition), the demixing threshold shifts to higher densities, suggesting that strongly polydisperse systems are unlikely to exhibit this transition. 
The bidisperse system studied here therefore provides the clearest case where the clustering and demixing instabilities occur close to each other.
Consistent with this picture, our simulations show that beyond $\rho_d$ the cluster crystal does not undergo macroscopic phase separation but instead develops predominantly two types of clusters with different local compositions, as shown in Fig.~\ref{fig:fig_3}\textbf{(a) (Top)}. 
We also observe that the distributions of local compositions corresponding to $A$-rich and $B$-rich clusters are neither identical nor symmetric about the global composition. 
This asymmetry originates from the difference in the average occupancies of the two types of clusters. 
Such behaviour arises naturally from the free energy minimisation of Eq.~\eqref{eq:F} when the local occupancies are constrained to fixed values. 
The resulting free energy landscape shows that $F$ is minimized for different local occupancies of $A$-rich and $B$-rich clusters, as illustrated in Fig.~\ref{fig:fig_3}\textbf{(a) (Bottom)}. 
This simplified approach qualitatively captures both the demixing transition point and the average local occupancies of the demixed clusters. 
As shown in Fig.~\ref{fig:fig_3}\textbf{(c)}, for $\rho\le\rho_d$ the orientational order parameters increase toward those of a perfect crystal with increasing density, whereas beyond the transition they decrease sharply, indicating an order-disorder transition. 
Physically, before the transition all lattice sites are equivalent and the system behaves as an effectively monodisperse crystal with $N_c$ particles whose energy scales as $\overline{n}\epsilon$. 
However, beyond $\rho_d$ the lattice sites differ in both total occupancy and composition, producing an effective bidisperse,  substitutionally disordered crystal whose increasing heterogeneity reduces structural order as the density increases.

We have shown that multicomponent ultrasoft particles can form cluster crystals that undergo a density driven transition from a homogeneous mixed state to a heterogeneous ``alloy'' with positionally ordering on the original lattice sites.   
This transition arises from the spontaneous differentiation of lattice sites into populations with distinct compositions and occupancies without macroscopic phase separation, revealing a mechanism for symmetry breaking in cluster forming systems  that is unique for mixtures. 
The demixing density follows a simple composition scaling predicted by the stability analysis of the homogeneous fluid, which captures the dependence of the transition on temperature, particle size mismatch, and global composition. 
This scaling suggests that heterogeneous cluster solids are most readily stabilized at low temperatures and near equal component concentrations. 
Similar behavior is observed for both FCC and BCC cluster crystals, with the transition occurring at slightly lower densities for the BCC lattice while the free energies of the two structures remain nearly identical. 
Our results further indicate that mixtures with identical particle sizes but asymmetric interactions provide a simpler realization of the same instability, where packing effects are absent: clusters demix when the cross interaction exceeds the average self interaction, $\epsilon_{AB}>(\epsilon_{AA}+\epsilon_{BB})/2$, while the demixing density diverges for symmetric interactions. 
In this case an interaction asymmetry parameter $\eta_2=\epsilon_{AB}/\langle\epsilon\rangle$ plays a role analogous to the size ratio in controlling the transition. 
For $\epsilon_{AA}=\epsilon_{BB}\ne\epsilon_{AB}$ the system is expected to develop compositional heterogeneity without occupancy heterogeneity, leading to two interpenetrating but equivalent $A$- and $B$-rich sublattices, analogous to ordering transitions in binary alloys such as CuAu at equimolar composition. 
Beyond the demixing transition the structural order gradually decreases with density, suggesting that the system may eventually undergo an amorphization transition distinct from that observed in conventional soft particle solids~\cite{tong2015crystals}. 
Compositional heterogeneity may therefore strongly influence collective excitations, potentially modifying the low frequency vibrational spectrum as the system evolves from a cluster crystal toward an amorphous state~\cite{schirmacher2024nature,xu2024low}.

\begin{acknowledgments}
This work was supported by the International Scientists Project funded by the Beijing Natural Science Foundation (Grant No. IS25041). We acknowledge useful discussions with Robert Jack, Michiel Sprik, and Kabir Ramola. Computational resources were provided by the Institute of Physics, Chinese Academy of Sciences, and the Yusuf Hamied Department of Chemistry, University of Cambridge.
\end{acknowledgments}

\appendix

\section{Quantifying compositional heterogeneity}\label{app:nafnbf}
\begin{figure}[h!]
\centering
\includegraphics[width=0.9\columnwidth]{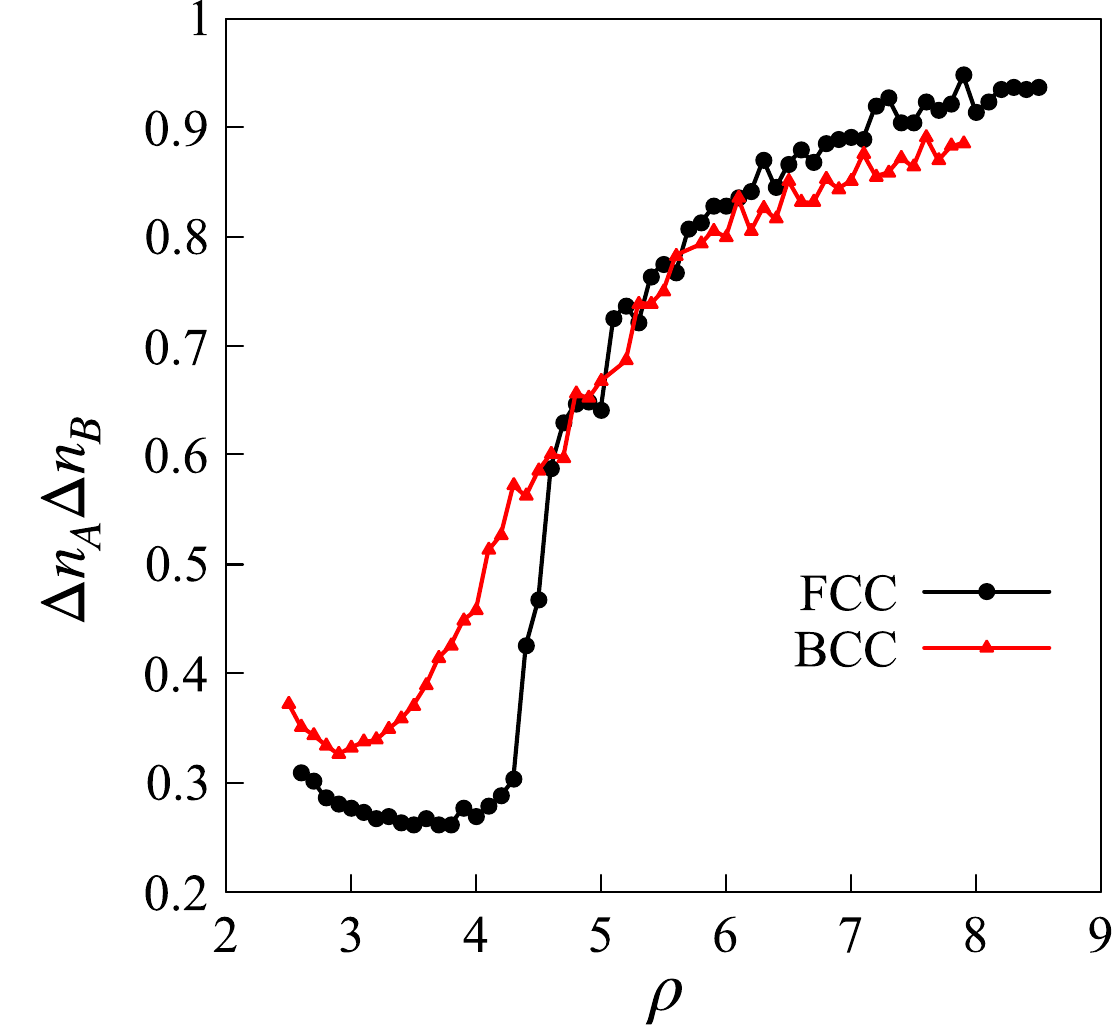}
\caption{\label{fig:fig_4}
Compositional Heterogeneity in BCC and FCC cluster crystal as a function of density at $T=0.3,\eta=1.2$.
}
\end{figure}
To quantify the homogeneity of the cluster crystal we analyze fluctuations in the lattice site occupancies. 
Let $n_A^i$ and $n_B^i$ denote the number of $A$ and $B$ particles occupying lattice site $i$, and let $n^i=n_A^i+n_B^i$ be the total occupancy of that site. The average occupancies are defined as
\begin{equation}
\bar n_\mu = \frac{1}{N_c}\sum_{i=1}^{N_c} n_\mu^i , \qquad
\bar n = \frac{1}{N_c}\sum_{i=1}^{N_c} n^i ,
\end{equation}
where $\mu\in\{A,B\}$ and $N_c$ is the total number of lattice sites.
Fluctuations in the occupancies are quantified by the normalized variances
\begin{equation}
\Delta n_\mu = \frac{\sqrt{\langle \overline{(n_\mu^i-\bar n_\mu)^2} \rangle}}{\bar n_\mu},
\qquad
\Delta n = \frac{\sqrt{\langle \overline{(n^i-\bar n)^2} \rangle}}{\bar n},
\end{equation}
where $\langle \cdot \rangle$ denotes configurational averaging.
To characterize compositional heterogeneity we consider the product of the species resolved fluctuations,
$\Delta n_A \Delta n_B$.
This quantity provides a convenient measure of compositional differentiation between clusters. For a perfectly homogeneous mixed crystal, where each lattice site contains particles in the same proportion as the global composition, the fluctuations vanish and $\Delta n_A \Delta n_B=0$. 
In the other extreme, if we only consider $N_C^A$ sites have only $A$ and $N_C^B$ sites only $B$ particles, then $\Delta n_A =\sqrt{N_c^B/N_c^A} $, and $\Delta n_B =\sqrt{N_c^A/N_c^B} $. So, in the case of complete demixing, $\Delta n_A\Delta n_B =1$.  

\section{Free energy of bidisperse cluster
crystals}\label{app:DFT_free}

Within a mean field approximation, the Helmholtz free energy functional of a
binary mixture reads
\begin{equation}
F[\{\rho_\mu\}]
=
F_{\rm id}+F_{\rm ex},
\end{equation}
with
\begin{align}
F_{\rm id}
&=
k_BT\sum_{\mu=A,B}
\int d^3r\,\rho_\mu(\mathbf r)
\left[
\ln(\rho_\mu(\mathbf r)\Lambda_\mu^3)-1
\right],\label{eq:Fid} \\
F_{\rm ex}
&=
\frac12\sum_{\mu\nu}
\int d^3r\,d^3r'\,
\rho_\mu(\mathbf r)\rho_\nu(\mathbf r')
\phi_{\mu\nu}(|\mathbf r-\mathbf r'|).
\label{eq;Fex}
\end{align}

Here $\rho_\mu(\mathbf r)$ denotes the one body density of species $\mu$,
$\Lambda_\mu$ is the thermal de Broglie wavelength, and
$\phi_{\mu\nu}(r)$ is the pair interaction potential. 
In the cluster crystal phase, particles localize around the sites
$\{\mathbf R\}$ of a Bravais lattice. We approximate the density profiles by a
sum of Gaussians,
\begin{equation}
\rho_\mu(\mathbf r)
=
\sum_{\mathbf R}
n_\mu (\textbf{R})
\left(\frac{\alpha_\mu}{\pi}\right)^{3/2}
\exp\!\left[-\alpha_\mu(\mathbf r-\mathbf R)^2\right],
\label{eq:rho_gaussian}
\end{equation}
where $n_\mu(\textbf{R})$ is the number of particles of species $\mu$ at the lattice site $\textbf{R}$ and
$\alpha_\mu$ controls the degree of localization. Using Eq.~\eqref{eq:rho_gaussian} in Eqs.~\eqref{eq:Fid} and ~\eqref{eq;Fex}, we obtain the follwing expression for per particle free energy,

\begin{equation}
    \begin{aligned}
        \frac{F}{N}&
= 
\frac{1}{nN_c}\left[
\underbrace{k_BT\sum_{i=1}^{N_c}\sum_{\mu}
n_i^{\mu}(\ln \left(n_i^{\mu}C_{\mu})-5/2\right)}_{\textbf{Ideal}}\right.\\
&+
\left.\underbrace{\sum_i
\sum_{\mu,\nu}
\Big[
n_i^{\mu}n_i^{\nu}
-\delta_{\mu\nu}n_i^{\mu}
\Big]
U_{\mu\nu}}_{\textbf{Intra-Cluster}}
+
\underbrace{\sum_{i\neq j}
\sum_{\mu,\nu}
n_i^{\mu}n_j^{\nu}
V_{\mu\nu}(R_{ij})}_{\textbf{Inter-Cluster}},\right]
    \end{aligned}
\end{equation}
where, $C_{\mu} = (\lambda^2_{\mu}\alpha_{\mu}/\pi)^{3/2}$ and the intracluster and intercluster interaction kernels are given by
\begin{align}
U_{\mu\nu}
=&
\sqrt{\frac{\alpha_{\mu\nu}^{3}}{2\pi}}
\int_0^\infty dr\, r^2
e^{-\alpha_{\mu\nu}r^2/2}
\phi_{\mu\nu}(r),
\\
V_{\mu\nu}
=&
\sqrt{\frac{\alpha_{\mu\nu}}{8\pi }}
\int_0^\infty dr\, \frac{r}{R}
\phi_{\mu\nu}(r)\\
&\hspace{1.3cm}\left(
e^{-\alpha_{\mu\nu}(r-R)^2/2}
-
e^{-\alpha_{\mu\nu}(r+R)^2/2}
\right).
\end{align}

\section{Stability analysis for multicomponent systems}

To generalize the demixing criterion to a multicomponent system, we consider small density fluctuations around a homogeneous reference state, $\rho_\mu(\mathbf r)=\rho_\mu^0+\delta\rho_\mu(\mathbf r)$, where $\mu=1,\dots,M$ labels the species. Expanding the free energy functional to quadratic order in the fluctuations yields
\begin{equation}
\delta^2F=\frac{1}{2}\sum_{\mu\nu}\int d\mathbf r\, d\mathbf r'\,
\delta\rho_\mu(\mathbf r)\,
H_{\mu\nu}(|\mathbf r-\mathbf r'|)\,
\delta\rho_\nu(\mathbf r'),
\end{equation}
where the kernel is
\begin{equation}
H_{\mu\nu}(r)=\frac{k_BT}{\rho_\mu}\delta_{\mu\nu}\delta(r)+\phi_{\mu\nu}(r).
\end{equation}

Fourier transforming the fluctuations, $\delta\rho_\mu(\mathbf r)=\sum_{\mathbf q}\delta\rho_\mu(\mathbf q)e^{i\mathbf q\cdot\mathbf r}$, gives
\begin{equation}
\delta^2F=\frac{V}{2}\sum_{\mathbf q}\sum_{\mu\nu}
\delta\rho_\mu(\mathbf q)\,
H_{\mu\nu}(\mathbf q)\,
\delta\rho_\nu(-\mathbf q),
\end{equation}
with the stability matrix
\begin{equation}
H_{\mu\nu}(q)=\frac{k_BT}{\rho_\mu}\delta_{\mu\nu}+\tilde{\phi}_{\mu\nu}(q),
\end{equation}
where $\tilde{\phi}_{\mu\nu}(q)$ is the Fourier transform of the interaction potential.
Here $\rho_\mu=x_\mu\rho$ and $x_\mu$ are the mole fractions. The eigenvalues follow from the equation $\sum_\nu H_{\mu\nu}(q)v_\nu=\lambda v_\mu$. One eigenvector corresponds to a total density fluctuation, $v_\mu\propto x_\mu$, for which all species fluctuate proportionally and the composition remains unchanged. Substituting $v_\mu=x_\mu$ into the eigenvalue equation and summing over $\mu$ gives the density eigenvalue
\begin{equation}
\lambda_{\mathrm{dens}}(q)=\frac{k_BT}{\rho}+\sum_{\mu\nu}x_\mu x_\nu \tilde{\phi}_{\mu\nu}(q).
\end{equation}

The remaining eigenmodes correspond to concentration fluctuations that conserve the total density and therefore satisfy $\sum_\mu v_\mu=0$. This constraint removes one degree of freedom from the $M$-component space, leaving $M-1$ independent concentration modes. A convenient basis is given by pair exchange vectors with components $v_\mu=1$, $v_\nu=-1$, and zero otherwise. Projecting the eigenvalue equation onto such a mode yields

\begin{equation}
\lambda_{\mu\nu}(q)=
\frac{k_BT}{\rho}\left(\frac{1}{x_\mu}+\frac{1}{x_\nu}\right)
+\tilde{\phi}_{\mu\mu}(q)
+\tilde{\phi}_{\nu\nu}(q)
-2\tilde{\phi}_{\mu\nu}(q).
\end{equation}

Although there are $M(M-1)/2$ such pair directions, they are not linearly independent due to the constraint $\sum_\mu v_\mu=0$, and therefore only $M-1$ independent demixing eigenvalues exist. The homogeneous phase becomes unstable when the smallest eigenvalue of $\textbf{H}(q)$ vanishes.

\bibliography{cluster}

\end{document}